\documentclass[conference]{IEEEtran}
\usepackage{silence}
\ifCLASSINFOpdf
\else
\fi
\usepackage{braket}
\usepackage{amsmath}   % For advanced mathematical typesetting
\usepackage{amssymb}   % For additional mathematical symbols
\usepackage{amsthm}    % For theorem environments

\usepackage{algorithmic}
\usepackage{url}
\usepackage{graphicx}
\usepackage{caption}
\usepackage{subcaption}
\usepackage{color}

\usepackage{physics}
\usepackage{hyperref}
\usepackage{booktabs}
\usepackage{algorithm} % For using the algorithm environment
\usepackage{multirow} 

\usepackage{xcolor}
\hypersetup{breaklinks=true, colorlinks=true, linkcolor={blue!90!black}, citecolor={blue!90!black}, urlcolor={blue!90!black}}
\usepackage{doi}
\usepackage{amsmath,amssymb}

\begin{document}
\title{State Dependent Optimization with Quantum Circuit Cutting}
\author{
    \IEEEauthorblockN{
        Xinpeng Li\IEEEauthorrefmark{1},
        Ji Liu\IEEEauthorrefmark{2},
        Jeffrey M. Larson\IEEEauthorrefmark{2},
        Shuai Xu\IEEEauthorrefmark{1},
        Sundararaja Sitharama Iyengar\IEEEauthorrefmark{3},
        Paul Hovland\IEEEauthorrefmark{2},
        Vipin Chaudhary\IEEEauthorrefmark{1}
    }
    \IEEEauthorblockA{\IEEEauthorrefmark{1}Dept. of Computer and Data Science, Case Western Reserve University, Cleveland, OH\\ Email: \{xxl1337, ssx214, vxc204\}@case.edu}
    \IEEEauthorblockA{\IEEEauthorrefmark{2}Mathematics and Computer Science Division, Argonne National Laboratory, Lemont, IL\\ Email: \{ji.liu, jmlarson, hovland\}@anl.gov}
    \thanks{\IEEEauthorrefmark{1}Xinpeng Li and \IEEEauthorrefmark{2}Ji Liu contributed equally to this work}
    \IEEEauthorblockA{\IEEEauthorrefmark{3}Knight Foundation School of Computing and Information Sciences, Florida International University, Miami, FL\\ Email: \{iyengar\}@cs.fiu.edu}
}
% \author{
% \IEEEauthorblockN{
% \begin{minipage}{0.47\linewidth}
% \centering
% Xinpeng Li \\
% \textit{Dept. of Computer and Data Science} \\
% \textit{Case Western Reserve University} \\
% Cleveland, US \\
% xxl1337@case.edu
% \end{minipage}
% \hfill
% \begin{minipage}{0.47\linewidth}
% \centering
% Ji Liu \\
% \textit{Mathematics and Computer Science Division} \\
% \textit{Argonne National Laboratory} \\
% Lemont, US \\
% ji.liu@anl.gov
% \end{minipage}
% }

% \vspace{1em}

% \IEEEauthorblockN{
% \begin{minipage}{0.47\linewidth}
% \centering
% Jeffrey M. Larson \\
% \textit{Mathematics and Computer Science Division} \\
% \textit{Argonne National Laboratory} \\
% Lemont, US \\
% jmlarson@anl.gov
% \end{minipage}
% \hfill
% \begin{minipage}{0.47\linewidth}
% \centering
% Shuai Xu \\
% \textit{Dept. of Computer and Data Science} \\
% \textit{Case Western Reserve University} \\
% Cleveland, US \\
% ssx214@case.edu
% \end{minipage}
% }

% \vspace{1em}

% \IEEEauthorblockN{
% \begin{minipage}{0.47\linewidth}
% \centering
% Paul Hovland \\
% \textit{Mathematics and Computer Science Division} \\
% \textit{Argonne National Laboratory} \\
% Lemont, US \\
% hovland@mcs.anl.gov
% \end{minipage}
% \hfill
% \begin{minipage}{0.47\linewidth}
% \centering
% Vipin Chaudhary \\
% \textit{Dept. of Computer and Data Science} \\
% \textit{Case Western Reserve University} \\
% Cleveland, US \\
% vxc204@case.edu
% \end{minipage}
% }
% }


\maketitle
% \begingroup\renewcommand\thefootnote{\fnsymbol{footnote}}
% \footnotetext[4]{Equal contribution}
% \footnotetext[7]{This research was supported in part by NSF Awards 2216923, 2117439, 2238734, 2217021 and 2311950}
% \endgroup
\begin{abstract}
Quantum circuits can be reduced through optimization to better fit the constraints of quantum hardware. One such method, initial-state dependent optimization (ISDO), reduces gate count by leveraging knowledge of the input quantum states.

Surprisingly, we found that ISDO is broadly applicable to the downstream circuits produced by circuit cutting. Circuit cutting also requires measuring upstream qubits and has some flexibility of selection observables to do reconstruction. Therefore, we propose a state-dependent optimization  (SDO) framework that incorporates ISDO, our newly proposed measure-state dependent optimization (MSDO), and a biased observable selection strategy. Building on the strengths of the SDO framework and recognizing the scalability challenges of circuit cutting, we propose non-separate circuit cutting—a more flexible approach that enables optimizing gates without fully separating them.

We validate our methods on noisy simulations of QAOA, QFT, and BV circuits. Results show that our approach consistently mitigates noise and improves overall circuit performance, demonstrating its promise for enhancing quantum algorithm execution on near-term hardware.
\end{abstract}


% \fi
\IEEEpeerreviewmaketitle

\section{Introduction}
% Quantum computing is a novel paradigm with potential applications in cryptography~\cite{pirandola2020advances},
% % material science~\cite{georgescu2014quantum}, 
% optimization~\cite{moll2018quantum}, and artificial intelligence~\cite{huang2021power}, enabling solutions to problems intractable for classical computers. However, its practical use remains limited, largely due to increasing noise as circuit sizes grow. To mitigate this, 
Quantum circuit optimization~\cite{saeedi2013synthesis, nash2020quantum} focuses on generating more efficient and streamlined circuits to improve circuit fidelity. One such quantum circuit optimization approach is initial-state dependent optimizations (ISDO)~\cite{liu2021relaxed, jang2022initial}, which leverages knowledge of the quantum input states to eliminate or modify gates without altering the circuit's output. However, ISDO is limited in its applicability to the front few layers of the circuit.

% In this work, we identify that ISDO is well-suited for circuit cutting \cite{peng2020simulating}—a technique that enables the simulation of large quantum circuits on smaller devices. The core idea of circuit cutting is wire cutting: measuring upstream qubits in some observables and preparing the downstream qubit in various initial states. Surprisingly, Existed ISDO can frequently be applied on these downstream subcircuits, improving the output fidelity. For upstream subcircuits, we propose Measure-State Dependent Optimization (MSDO), which is similar than ISDO but doing optimization on measure. Becauase upstream and downstream always apear as pair in wire cutting, for conveinount, we propose ISDO on donswtream and MSDO on upstream as a State Dependent Optimization (SDO) frame work, a specifically designed circuit optimization framework for wire cutting. 

\begin{figure}[htbp]
    \centering
    \includegraphics[width=\linewidth]{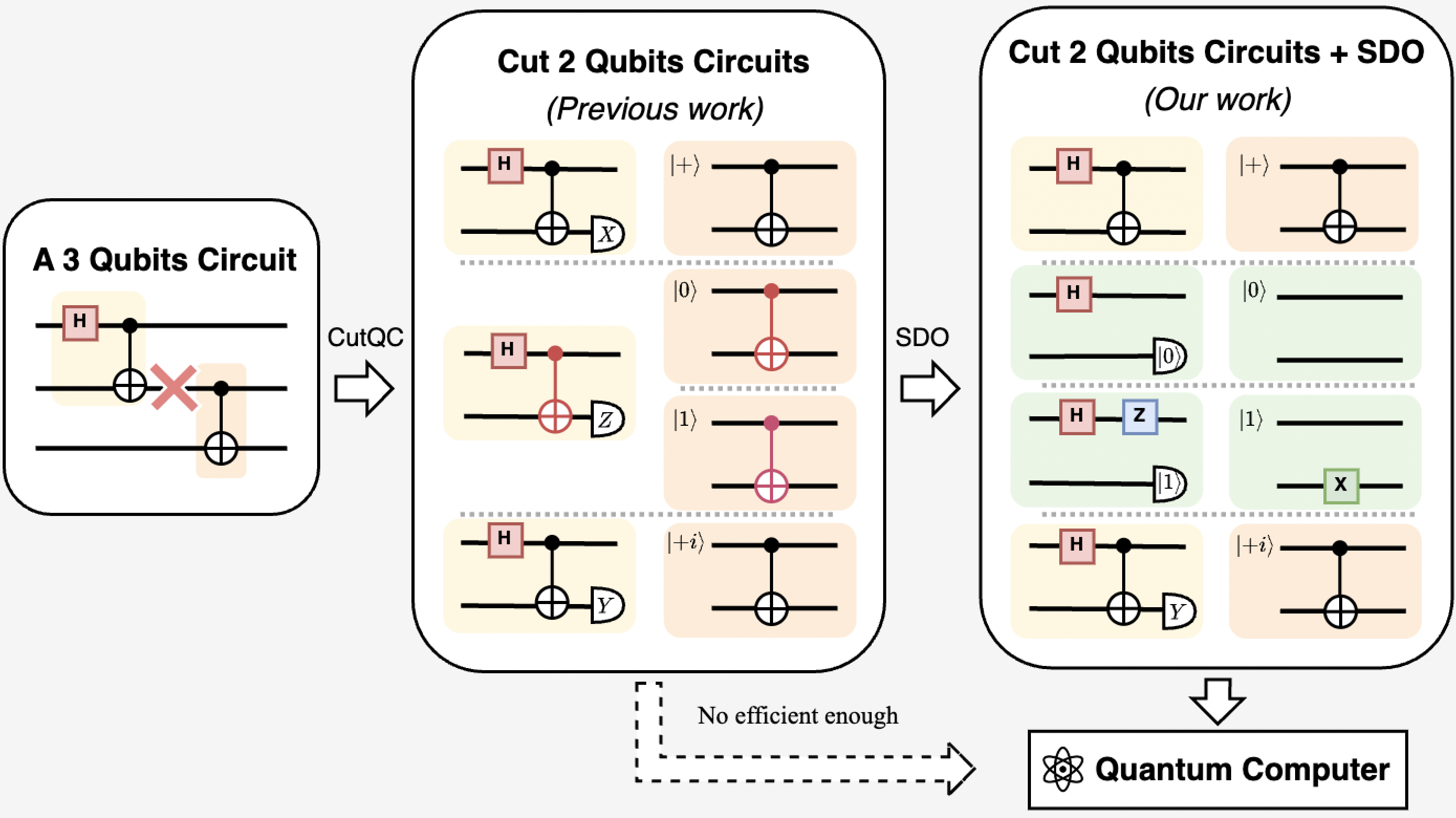} 
    \caption{Framework of circuit cutting with \textit{state-dependent optimization} (SDO): Assuming we need to run a three-qubit circuit on a two-qubit device, circuit cutting can be used to divide it into two-qubit subcircuits. The SDO framework can further reduce gates, making certain subcircuits easier to execute.}
    \label{fig: Circuit Cutting Framework}
\end{figure}

In this work we show that ISDO is ideally suited to circuit cutting \cite{peng2020simulating}, which enables the simulation of large quantum circuits on smaller devices. Circuit cutting primarily relies on wire cutting: measuring upstream qubits in various observables and preparing downstream qubits in various initial states. This process splits the original circuit into upstream and downstream subcircuits, each of which is smaller than the uncut circuit. We find that existing ISDO can often be applied to downstream subcircuits to improve output fidelity. For upstream subcircuits, we introduce measure‐state dependent optimization (MSDO), which mirrors ISDO but applies optimization at the measurement stage. Observing that different subcircuits exhibit varying fidelities, we introduce a biased observable selection strategy to choose those subcircuits that yield higher fidelity in wire‐cutting reconstruction. We then integrate downstream ISDO, upstream MSDO, and biased observable selection into a unified state‐dependent optimization (SDO) framework—a circuit optimization scheme specifically tailored for wire cutting. An overview of circuit cutting with the SDO framework is shown in Figure~\ref{fig: Circuit Cutting Framework}.

We observe that while the SDO framework improves the fidelity of results in circuit cutting, its applicability is limited because the number of subcircuits grows exponentially with the number of cuts. To address this situation, we extend circuit cutting to non-separate circuit cutting for optimization purpose. Non-separate circuit cutting does not require cuts to divide a circuit into fully separated fragments. This enables greater flexibility in choosing cutting locations and limiting the number of cuts. 
% The method offers a middle option between cutting the circuit and leaving it uncut.

% Finally, for both circuit cutting and Non-Separate Circuit Cutting, the identity \(I\) observable is always computed using subcircuits of other observables. Commonly, it use \(Z\) to calculate \(I\) \cite{tang2021cutqc}. However, 

% Finally, by selecting higher-fidelity subcircuits to calculate the identity \(I\) observable, we can further enhance the robustness of SDO. We call this approach ``Biased Observable Wire Cutting" and integrate it into the SDO.

The main contributions of this paper are as follows.
\begin{itemize}
    % \item We propose an MSDO, which can optimize circuit from measurement by runing them mult times.

    \item We propose an SDO framework that incorporates ISDO, MSDO, and biased observable selection for wire cutting.

    % \item We introduce a biased basis selection method for wire cutting to enhance the SDO framework.

    \item We propose non-separate circuit cutting for more flexible application of the SDO framework.
    % \item We provide an easy-to-use method for detecting optimized gates, which aids in selecting biased observables and determining appropriate cutting locations.
\end{itemize}

\section{Background and Motivation}
\label{sec:Background}

\subsection{Circuit Cutting}

Circuit cutting allows us to execute a large circuit by running smaller, more manageable subcircuits. To perform circuit cutting, three steps must be completed. First, we identify cut locations along the circuit’s wires; making cuts at these points decomposes the circuit into independent fragments. For example, the circuit shown on the rightmost side of Figure~\ref{fig: Circuit Cutting Framework} can be cut on the second qubit (marked by a red cross), splitting it into two separate fragments. Second, at each cut location, the upstream wire is measured in a chosen set of observables, and the downstream wire is initialized in the corresponding eigenstates. Consequently, each fragment must be executed multiple times; we refer to each circuit that needs to be executed as a subcircuit. A commonly used basis is the Pauli set \(\mathcal{B} = \{I, X, Y, Z\}\). More precisely, wire cutting decomposes a single-qubit density matrix \(\rho\) as
\begin{equation}
\rho = \frac{1}{2}\bigl( \mathbf{Tr}(I\rho)\,I +  \mathbf{Tr}(X\rho)\,X +  \mathbf{Tr}(Y\rho)\,Y +  \mathbf{Tr}(Z\rho)\,Z\bigr).
\label{eq:WireCutting}
\end{equation}
In this decomposition each term \( \mathbf{Tr}(M\rho)\,M\) represents measuring the upstream wire with operator \(M\) and preparing the downstream wire in the corresponding eigenstate \(\ket{u}\) or \(\ket{v}\), where
\(
M = r\,\ket{u}\bra{u} + s\,\ket{v}\bra{v}.
\)
Note that any single-qubit density matrix can be decomposed into projectors onto four orthonormal basis states. In CutQC~\cite{tang2021cutqc} the authors choose the basis
\(\{\ket{0},\ket{1},\ket{+},\ket{+i}\}\),
reducing the required initializations from six to four. We adopt this approach in this paper, as illustrated in the middle panel of Figure~\ref{fig: Circuit Cutting Framework}. We execute all constructed subcircuits and reconstruct the result according to Equation~\ref{eq:WireCutting}.

Circuit cutting introduces two major overheads: the classical reconstruction process and the number of quantum subcircuit executions required. The total number of required subcircuits and the number of classical reconstruction float operators scale exponentially with the number of cuts. Several works~\cite{li2024efficient, chen2023online, chen2023efficient} have proposed methods to mitigate these overheads.

\subsection{Initial-State Dependent Optimization }
\label{subsec: Initial-State Dependent Optimization}
ISDO is a quantum optimization method conditioned on the initial state. It was  introduced in the work on relaxed peephole optimization (RPO)~\cite{liu2021relaxed} and extended to a more general case for multicontrol qubit gates by \cite{jang2022initial}. It can be applied to different types of gates, such as single-qubit gates, CNOT gates, multiqubit gates, and swap gates. 

\begin{figure}[htbp]
    \centering
    \begin{subfigure}[b]{0.47\linewidth}
        \centering
        \includegraphics[width=\linewidth]{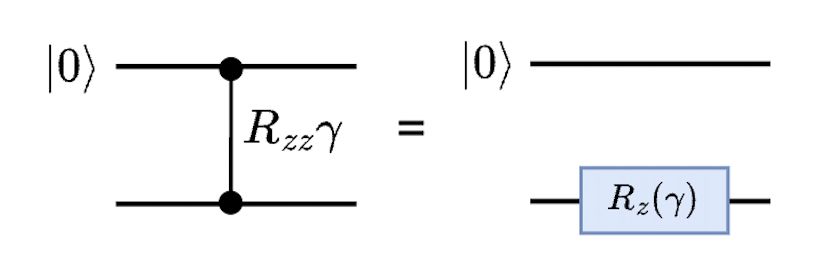}
        \caption{ISDO for RZZ with input qubit \(\ket{0}\)}
        \label{fig:figure2}
    \end{subfigure}
    \hfill
    \begin{subfigure}[b]{0.47\linewidth}
        \centering
        \includegraphics[width=\linewidth]{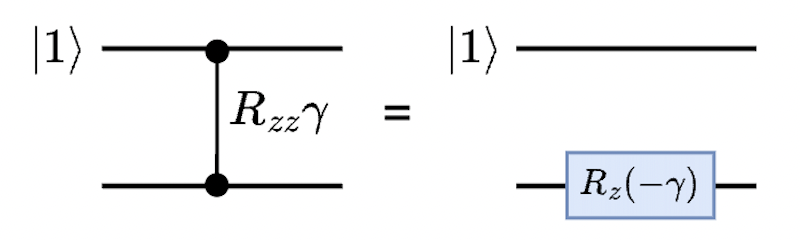}
        \caption{ISDO for RZZ with input qubit \(\ket{1}\)}
        \label{fig:figure3}
    \end{subfigure}
    \caption{Illustrations of ISDO on RZZ gate. For the RZZ gate, two input qubits are equivalent.}
    \label{fig: ISDO for RZZ}
\end{figure}

The simplest example is a single-qubit gate. We can omit the gate if the input state $\ket{\phi}$ matches the eigenstate of the unitary gate $U$ with an eigenvalue of 1, that is, $\ket{\phi}=U\ket{\phi}$. For a multiqubit gate, if we apply a global unitary $U$ between systems $A$ with a pure state \(\ket{\phi} \bra{\phi}_A\) and $B$ with density matrix \(\rho_B\), and $B$ remains independent of $A$ after the operation,

\begin{equation}
U \bigl(\ket{\phi} \bra{\phi}_A \otimes \rho_B \bigr)\,U^\dagger
= \ket{\phi} \bra{\phi}_A \otimes \rho'_B,
\label{eq: ISDO}
\end{equation}
then there exists  a local unitary $U_B$ to replace \(U\) such that
\[
\rho'_B = U_B\,\rho_B\,U_B^\dagger,
\]
and hence
\[
U \bigl(\ket{\phi} \bra{\phi}_A \otimes \rho_B \bigr)\,U^\dagger
= \ket{\phi} \bra{\phi}_A \otimes\
\bigl(U_B\,\rho_B,U_B^\dagger\bigr).
\]
When performing ISDO, we do not need to know the exact form of \(\rho_B\). A simple way to determine whether \(U\) can be replaced by \(U_B\) is to check whether
\begin{equation}
U \bigl(\ket{\phi}_A \otimes I_B\bigr)
= \ket{\phi}_A \otimes U_B,
\label{eq: ISDO unitary finding}
\end{equation}
in other words, if we can factor \(U\bigl(\ket{\phi}_A \otimes I_B\bigr)\) into \(\ket{\phi}_A\) tensored with a unitary on \(B\). Following  Equation \ref{eq: ISDO unitary finding}, we can check and optimize the gates from the initial state to the measurement until the initial state is not a pure state anymore. We can also prebuild an optimization gate template to avoid calculating \(U_B\) each time.
% \begin{enumerate}
%   \item If the circuit matches one of the template cases, we optimize it directly.
%   \item Otherwise, we apply Equation~\ref{eq: ISDO} to determine whether further optimization is possible.
% \end{enumerate} 
As an example, we illustrate the optimization of the RZZ gate in Figure~\ref{fig: ISDO for RZZ}. For further details of other gates, please refer to \cite{liu2021relaxed,jang2022initial}.

% \begin{figure}[htbp]
%     \centering
%     \begin{subfigure}[b]{0.47\linewidth}
%         \centering
%         \includegraphics[width=\linewidth]{Figures/ISDOCNOT0.png}
%         \caption{ISDO for CNOT with control qubit \(\ket{0}\)}
%         \label{fig:figure1}
%     \end{subfigure}
%     \hfill
%     \begin{subfigure}[b]{0.47\linewidth}
%         \centering
%         \includegraphics[width=\linewidth]{Figures/ISDOCNOT1.png}
%         \caption{ISDO for CNOT with control qubit \(\ket{1}\)}
%         \label{fig:figure2}
%     \end{subfigure}
%     \vspace{1em} % Add some space between rows
%     \begin{subfigure}[b]{0.47\linewidth}
%         \centering
%         \includegraphics[width=\linewidth]{Figures/ISDOCNOT+.png}
%         \caption{ISDO for CNOT with target qubit \(\ket{+}\)}
%         \label{fig:figure3}
%     \end{subfigure}
%     \hfill
%     \begin{subfigure}[b]{0.47\linewidth}
%         \centering
%         \includegraphics[width=\linewidth]{Figures/ISDOCNOT-.png}
%         \caption{ISDO for CNOT with target qubit \(\ket{-}\)}
%         \label{fig:figure4}
%     \end{subfigure}
%     \caption{Illustrations of ISDO on CNOT gate.}
%     \label{fig:ISDO for CNOT}
% \end{figure}

% \subsection{Motivation}

% When performing circuit cutting, we always initialize eigenstates from all Pauli observables, as they are particularly suited for specific optimizations. For example, in Figure~\ref{fig: Circuit Cutting Framework}, ISDO cannot be directly applied to the rightmost circuit. However, after cutting it into seven subcircuits, two of them can be optimized using ISDO. A further optimization can be achieved using MSDO; if we execute one additional subcircuit, both subcircuits can then eliminate the CNOT gate. Our tests show that ISDO and MSDO lead to improved fidelity.

% For example, in QAOA, cuts often occur before a series of RZZ gates. These RZZ gates can be optimized if the input states are \(\ket{0}\) and \(\ket{1}\). In subcircuits initialized with \(\ket{0}\) and \(\ket{1}\), the series of RZZ gates can be reduced to single-qubit gates. Similarly, for measurements, we can achieve optimization by retaining the results of a single eigenstate.

% As shown in Table \ref{tab: Circuit Cutting SDO}, if we cut at the second qubit, we prepare and run 8 two-qubits subcircuits to reconstruct the original three-qubits circuit. Compared to the original circuit cutting approach, the subcircuits initialized in the Z observable can be optimized, making the Z observable results more robust. We observe that this optimization for Z observables can be extended to other observables, leading to improved fidelity.

\section{State-Dependent Optimization Framework for Wire Cutting}
\label{sec: State Depend Optimization(SDO) for Wire Cutting}
This section introduces the SDO framework comprising ISDO (Section \ref{subsec: Initial-State Dependent Optimization}), MSDO, and biased observable selection. We will first cover MSDO and biased observable selection and then formalize the SDO framework.
% First, the framework applies MSDO and ISDO to each subcircuit across the full basis set. Next, it employs a biased basis selection to choose high‐fidelity subcircuits for reconstruction. This section is organized as follows: 
% We begin by introducing Measure‐State Dependent Optimization (MSDO), then propose a biased observable selection strategy, present the unified SDO framework, and conclude with a method for estimating its performance.

% In this section, we first extend the concept of ISDO to Measure-State Dependent Optimization (MSDO). We then introduce a State-Dependent Optimization (SDO) framework that incorporates both ISDO and MSDO for wire cutting. Finally, we propose a biased observable selection strategy to enhance the SDO framework.
% Finally, we present some techniques for identifying optimized gates given a specific state.
\subsection{Measure-State Dependent Optimization }
\label{subsec:Measure State Depend Optimization}

ISDO can reduce gates based on a specific initial state as input. This action is possible because the initial state is known with certainty, allowing for simplifications in the circuit. For a measurement, however, the output state is uncertain because it might give  multiple possible states with some probability.%, so it is not a certain state. 
% Deferred Measurement Principle works is because it maintains uncertainty by using classical control unitary gate. However, due to current techniques, classical control could introduce more noise than quantum gates; thus, we want to avoid it and also optimize gates just like Deferred Measurement Principle.
Therefore, for MSDO, we perform two steps: (1) convert the measured uncertain state into a definite state and (2) carry out the optimization.

\textbf{Step 1.}   Suppose we are measuring the first qubit of a state \(\rho\) with the observable $M$, where $M = rM_r + sM_s$ and $M_r and M_s$ are two projectors corresponding to eigenvalues \(r\) and \(s\), respectively. The \(n\)-qubit state \(\rho\) after measuring the first qubit is given by
\begin{equation}
\mathbf{Tr}_{1}\bigl(M\otimes I \cdot \rho\bigr)
= r\,\mathbf{Tr}_{1}\bigl(M_{r}\otimes I \cdot \rho\bigr)
+ s\,\mathbf{Tr}_{1}\bigl(M_{s}\otimes I \cdot \rho\bigr) .
\label{eq: MSDO executed twice}
\end{equation}
Based on Equation~\ref{eq: MSDO executed twice}, the circuit can be easily executed twice. The first time, we  keep only those outcomes corresponding to \( M_r \), and  we obtain the outcome of the term \( \mathbf{Tr}_1(M \otimes I \cdot  \rho) \). The second time, we  keep only those outcomes corresponding to \( M_s\), and we obtain the outcome of term \( \mathbf{Tr}_1(M_s \otimes I \cdot \rho) \). We can completely reconstruct the outcome of measuring on \(M\) by doing \( r\mathbf{Tr}_1(M_r \otimes I \cdot  \rho) + s\mathbf{Tr}_1(M_s \otimes I \cdot  \rho) \). This process transforms the measurement into a certain state.

\textbf{Step 2.} After completing Step 1, we optimize the circuits using a procedure analogous to ISDO. The optimization is defined as follows.

\begin{figure}[htbp]
    \centering
    \includegraphics[width=0.7\linewidth]{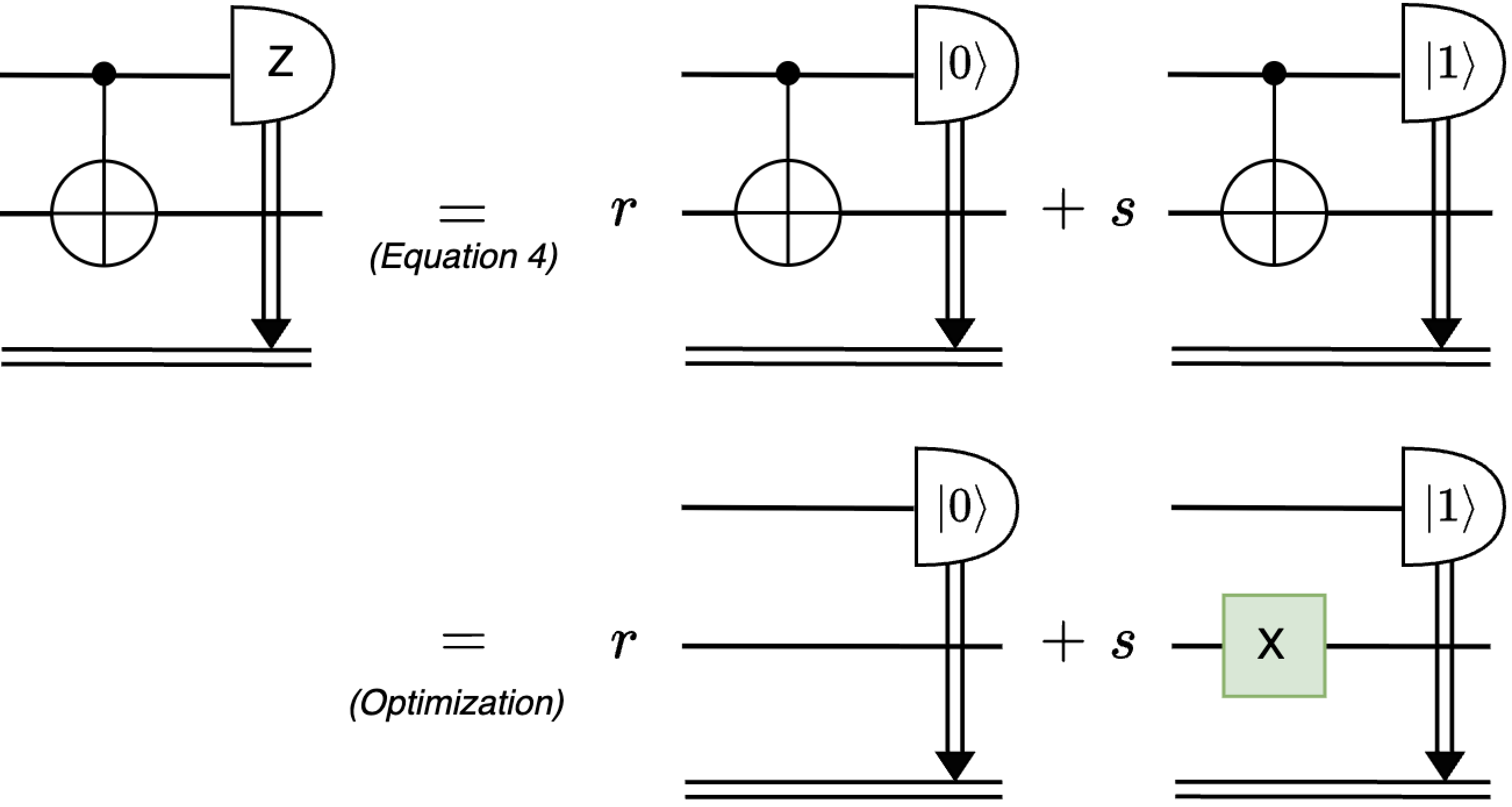}
    \caption{Example of \textit{measure-state dependent optimization} (MSDO). Measuring the first qubit in the Z observable is equivalent to measuring in the Z observable while separately extracting the results corresponding to \( \ket{0} \) and \( \ket{1} \) using two circuits. Then, by applying the inverse of ISDO, we obtain two circuits with fewer gates.}
    \label{fig:MSDO} 
\end{figure}

Consider the partial‐trace measurement
\(
\mathbf{Tr}_{1}\bigl[(M_r \otimes I)\,\rho\bigr].
\)
Suppose that immediately before this measurement we apply a unitary \(U\) to a pre‐unitary state \(\bar{\rho}\), so that
\(
\rho = U\,\bar{\rho}\,U^\dagger.
\)
We decompose \(\rho\) and \(\bar{\rho}\) as
\[
\rho
= \alpha\,M_r \otimes \rho_r
+ \beta\,M_s \otimes \rho_s,
\quad
\bar{\rho}
= \alpha'\,M_r \otimes \bar{\rho}_r
+ \beta'\,M_s \otimes \bar{\rho}_s.
\]
Since
\[
 \mathbf{Tr}_{1}\bigl[(M_r \otimes I)\,\rho\bigr]
=
 \mathbf{Tr}\bigl[(M_r \otimes I)\,\rho\bigr]\,\rho_r,
\]
if
\begin{equation}
 \mathbf{Tr}\bigl[(M_r \otimes I)\,\rho\bigr]\,\rho_r
=
 \mathbf{Tr}\bigl[(M_r \otimes I)\,\bar{\rho}\bigr]\,\rho_r,
\label{eq:MSDO}
\end{equation}
then \(U\) may be replaced by \(U_B\). It follows that
\[
U_B\,\bar{\rho}_r\,U_B^\dagger = \rho_r.
\]
As in Section~\ref{subsec: Initial-State Dependent Optimization}, we do not need to know the explicit forms of \(\rho_r\) or \(\bar{\rho}_r\) to perform optimization or determine \(U_B\). Instead, we check whether \(U\) can be optimized, and we extract \(U_B\) via
\begin{equation}
(\bra{u}\otimes I)\,U
= \bra{u}\otimes U_B,
\label{eq: MSDO unitary finding}
\end{equation}
where \(\bra{u}\) following \(M_r = \ket{u} \bra{u}\).
The only difference with ISDO is the direction of the check: MSDO performs it from the measurement back to the initial state. We provide an example of two steps of MSDO in Figure~\ref{fig:MSDO}.

% This optimization is different from ISDO: for ISDO, detection proceeds from front to back, whereas for MSDO, it proceeds from back to front. 
% We provide an example in Figure~\ref{fig:MSDO}.

A drawback of MSDO is that it requires running the circuit twice for one cut. If applied to \(k\) cuts, the number of circuits to execute grows exponentially as \(2^k\). However, considering that each subcircuit originally requires at least \(3^k\) executions, the increase to \(4^k\) represents only a polynomial overhead. When classical communication is relatively inexpensive, classically controlled single-qubit gates can be used to eliminate this overhead \cite{harada2024doubly}.

\subsection{Method for Estimating ISDO and MSDO Effectiveness}
\label{subsec: Evaluation of the Number of Optimized Gates}
We propose a simple method for assessing how many gates can be optimized for ISDO and MSDO.
Assume we want to determine whether an \(n\)-qubit unitary operator \( U \) can be optimized for a specific state \( \ket{\phi} \) on certain qubits. We can reorder the qubits so that qubits \(1\) to \(j\) are in unknown states and qubits \(j+1\) to \(n\) are in the specific pure state \( \ket{\phi} \). This optimization can be easily verified by checking whether \( U \) commutes with the state density matrix \( \ket{\phi}\bra{\phi} \) extended with an identity matrix, represented as \( I \otimes \ket{\phi}\bra{\phi} \). Mathematically, the commutation condition can be written as
\begin{equation}
U \bigl(I \otimes \ket{\phi}\bra{\phi}\bigr)
= \bigl(I \otimes \ket{\phi}\bra{\phi}\bigr)\,U .
\label{eq:commuting-check}
\end{equation}

For ISDO, substituting $\rho_B = I$ into Equation~\eqref{eq: ISDO} yields
\begin{equation}
  U\bigl(\ket{\phi}\bra{\phi}\otimes I\bigr)\,U^\dagger
  = \ket{\phi}\bra{\phi}\otimes I.
\end{equation}

For MSDO, applying Equation~\eqref{eq:MSDO} to an arbitrary state \(\rho\) yields
\begin{equation}
   \mathbf{Tr}\bigl[(M_r \otimes I)\,\overline{\rho}\bigr]
  =  \mathbf{Tr}\bigl[U^\dagger\,(M_r \otimes I)\,U\,\rho\bigr].
\end{equation}
Hence, at the operator level,
\begin{equation}
  U\,(M_r \otimes I)
  = (M_r \otimes I)\,U.
\end{equation}

We check that two-matrix commute involves only matrix multiplications; moreover, if \(U\) is well characterized, one can use its eigenvectors to confirm the commutation relation. Therefore, a commutation check is generally simpler than verifying the conditions in Equations~\ref{eq: ISDO unitary finding} and~\ref{eq: MSDO unitary finding}.

\subsection{Biased Observable Selection}
\label{subsec: Biased Observable Selection for Non-Separate Circuit Cutting}
During reconstruction, we need to choose an observable \(M\) to compute
\(
   \mathbf{Tr}(I\,\rho) =  \mathbf{Tr}(M_r\,\rho) +  \mathbf{Tr}(M_s\,\rho).
\)
For downstream initialization, the two eigenstates of \(M\) are used to form an orthonormal basis. For example, CutQC~\cite{tang2021cutqc} uses measurements in the \(Z\) basis to evaluate the identity term,
\(
   \mathbf{Tr}\bigl(\ket{0}\bra{0}\,\rho\bigr)
  +  \mathbf{Tr}\bigl(\ket{1}\bra{1}\,\rho\bigr),
\)
on the upstream qubit. For the downstream qubit, it combines the Z-eigenstates \(\{\ket{0},\ket{1}\}\) with the states \(\{\ket{+},\ket{+i}\}\) to form the orthonormal basis
\(\{\ket{0},\ket{1},\ket{+},\ket{+i}\}\).

Because subcircuits associated with the \(M\) have  more impact on the final reconstruction result, we call \(M\) the \emph{biased observable}. We select \(M\) by estimating the gate reduction achieved (see Section~\ref{subsec: Evaluation of the Number of Optimized Gates}), thereby enhancing overall reconstruction fidelity.

\subsection{SDO Framework}
The SDO framework begins by estimating the performance of ISDO and MSDO to guide cutting location selection and biased observable selection. We then apply ISDO and MSDO to each resulting subcircuit.

We further demonstrate that this framework is effective in most wire-cutting scenarios. For most commonly used gates, ISDO works on at least one eigenstate from each (\(X\), \(Y\), and \(Z\)), and MSDO works on at least one (\(X\), \(Y\), and \(Z\)) observable. For instance, on the \texttt{ibm\_fe} quantum device, operational gates include CZ, ID, RX, RZ, RZZ, SX, and X. The CZ gate is compatible with ISDO when the input qubits are in \(\ket{0}\) or \(\ket{1}\), and with MSDO in measuring in \(Z\) observables for both control and target qubits. 

Table \ref{tab: Circuit Cutting SDO} provides an example of applying the SDO framework to a cut QAOA circuit. The RZZ gates can be optimized into a  single-qubit gate in half of the subcircuits. 
% Among the four reconstruction terms, all four benefit from SDO. This enhancement improves the fidelity of the subcircuits and further boosts the overall reconstruction fidelity. In the following, we introduce \textit{Biased Observable Selection for Circuit Cutting}.

\begin{table}[htp]
\captionsetup{font=small}
\centering
\renewcommand{\arraystretch}{1.4}
\setlength{\tabcolsep}{6pt}
\begin{tabular}{c c}
\toprule
\multicolumn{2}{c}{\textbf{Uncut QAOA Circuit}} \\
\midrule
\multicolumn{2}{c}{\includegraphics[width=0.4\linewidth]{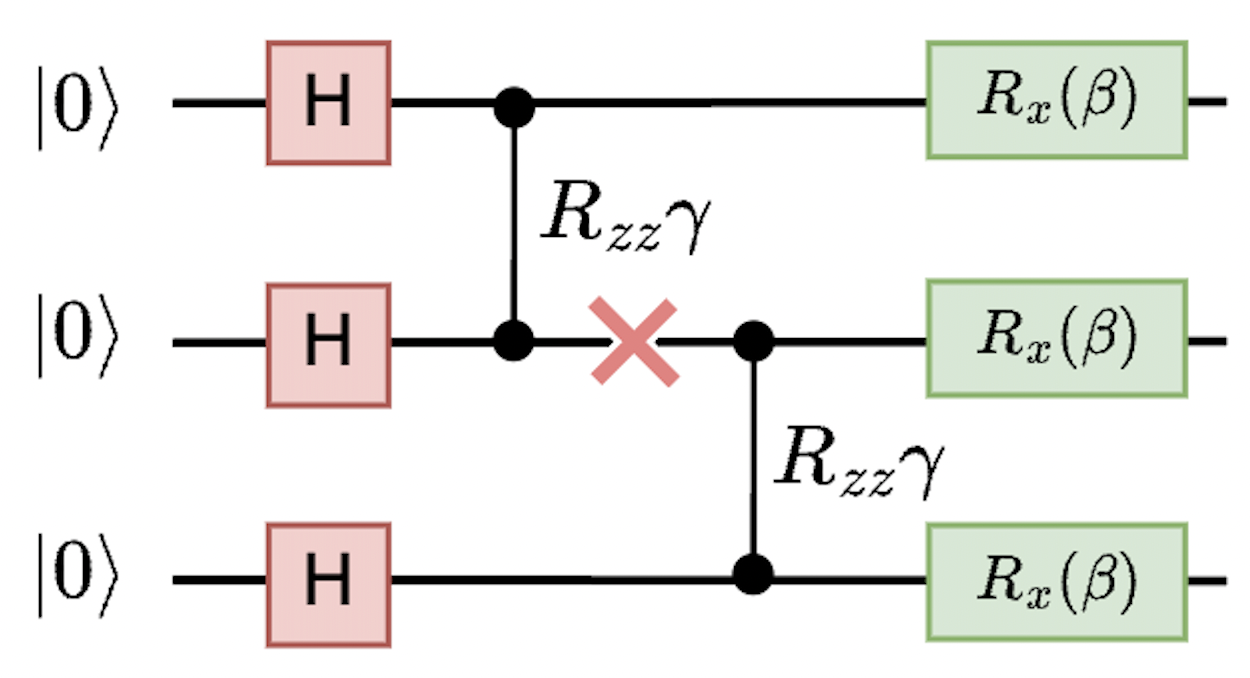}} \\
\midrule
\textbf{Upstream} & \textbf{Downstream} \\
\midrule
\textbf{Z Measurement $\ket{0}$} & \textbf{Z Initialization $\ket{0}$} \\
\includegraphics[width=0.4\linewidth]{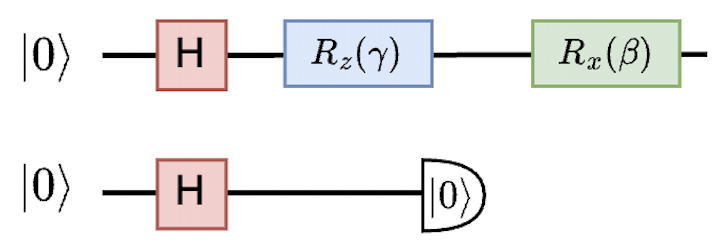} & 
\includegraphics[width=0.4\linewidth]{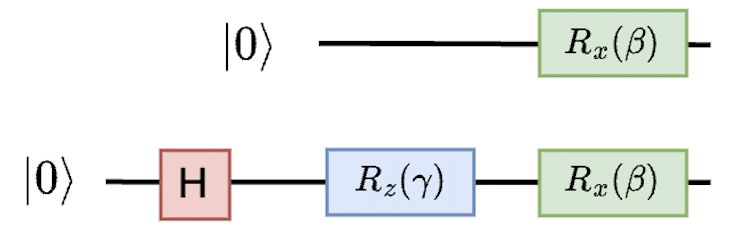} \\
\midrule
\textbf{Z Measurement $\ket{1}$} & \textbf{Z Initialization $\ket{1}$} \\
\includegraphics[width=0.4\linewidth]{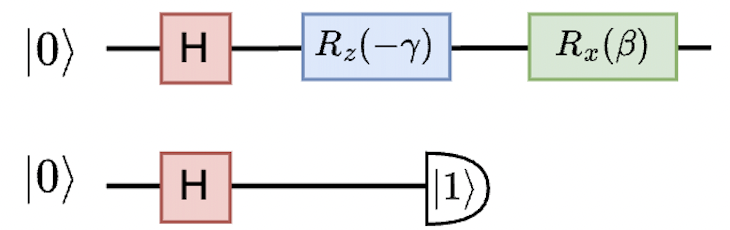} &
\includegraphics[width=0.4\linewidth]{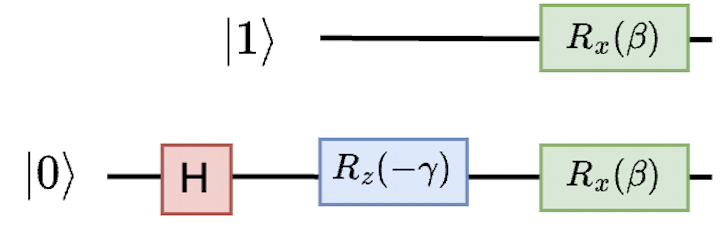} \\
\midrule
\textbf{X Measurement} & \textbf{X Initialization $\ket{+}$} \\
\includegraphics[width=0.4\linewidth]{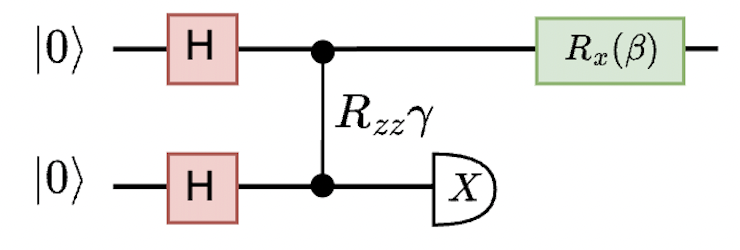} & 
\includegraphics[width=0.4\linewidth]{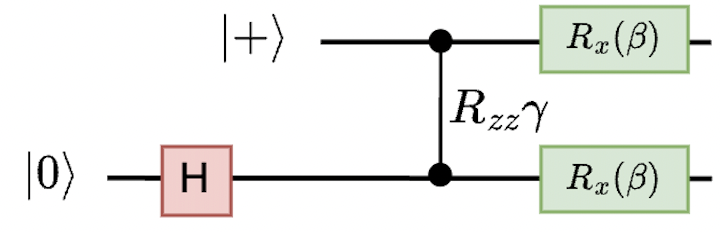} \\
\midrule
\textbf{Y Measurement} & \textbf{Y Initialization $\ket{+i}$} \\
\includegraphics[width=0.4\linewidth]{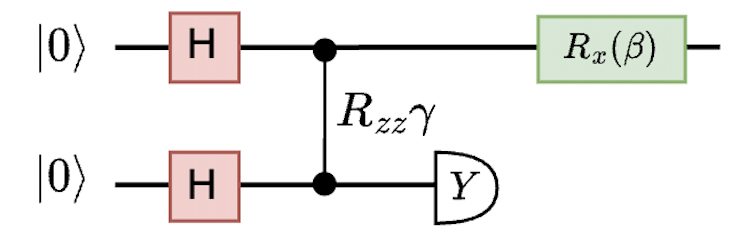} &
\includegraphics[width=0.4\linewidth]{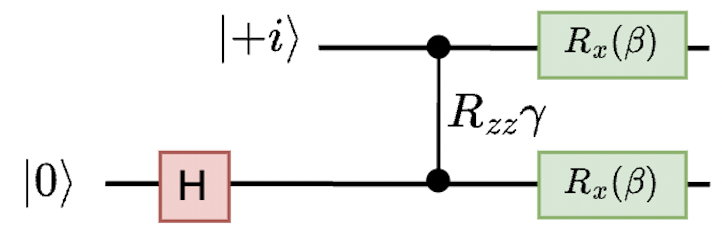} \\
\bottomrule
\end{tabular}
\caption{Example of applying circuit cutting with SDO on QAOA. The top shows the uncut QAOA circuit, and below are the eight subcircuits.}
\label{tab: Circuit Cutting SDO}
\end{table}

\section{flexiable Non-Seperate Circuit Cutting for SDO}

\label{sec: flexiable Non-seperate Circuit Cutting for SDO}

We observe that circuit cutting techniques have an inherent drawback: scalability. Introducing multiple cuts leads to an exponential increase in the number of subcircuits to execute, resulting in significant overhead for large circuits. However, we find that the SDO framework improves the fidelity of the reconstructed results compared with the uncut circuits. Therefore, we employ wire cutting solely for optimization purposes. In this section we first define non-separate circuit cutting and then illustrate how the SDO framework works with non-separate circuit cutting.

% \begin{table*}[ht]
% \centering
% \caption{Example of Circuit Cutting-Based Optimization}
% \begin{tabular}{|c|c|}
% \hline
% \multicolumn{2}{|c|}{Uncut QAOA Circuit} \\ 
% \hline
% \multicolumn{2}{|c|}{\includegraphics[width=0.3\linewidth]{Figures/QAOA_NSC_Uncut.png}} \\ 
% \hline
% Z Measurement \(\ket{0}\) & Z Measurement \(\ket{1}\) \\ 
% \hline
% \includegraphics[width=0.3\linewidth]{Figures/QAOA_NSC_Z0.png} & 
% \includegraphics[width=0.3\linewidth]{Figures/QAOA_NSC_Z1.png} \\ 
% \hline
% X Initialization \(\ket{+},\ket{-}\) & Y Initialization \(\ket{+i},\ket{-i}\) \\ 
% \hline
% \includegraphics[width=0.3\linewidth]{Figures/QAOA_NSC_X.png} & 
% \includegraphics[width=0.3\linewidth]{Figures/QAOA_NSC_Y.png} \\ 
% \hline
% \end{tabular}
% \end{table*}

\subsection{Non-Separate Circuit Cutting }
\label{subsec:Non-Separate Wire Cutting}
Non-separate circuit cutting (NSCC) also leverages the wire-cutting technique.
Unlike circuit cutting, however, non-separable cutting maintains a connection between the upstream and downstream segments, meaning the upstream term \(\mathbf{Tr}(M\rho)\) and the downstream term \(M\) cannot be decoupled. We provide an example in the top panel of Table \ref{tab: Non-Seperate Cutting}. We can see that a cut at the red cross mark does not separate the circuit; the upstream and downstream parts remain connected. Unlike the wire-cutting decomposition in circuit cutting shown in Equation~\ref{eq:WireCutting}, the most efficient qubit decomposition is given by
\begin{align}
\rho = & \, \mathbf{Tr}(\rho I) \cdot I + \mathbf{Tr}(\rho Z) \cdot \ket{0}\bra{0} - \mathbf{Tr}(\rho Z) \cdot \ket{1}\bra{1} \nonumber \\
& + \mathbf{Tr}(\rho X) \cdot \ket{+}\bra{+} - \mathbf{Tr}(\rho X) \cdot \ket{-}\bra{-} \nonumber \\
& + \mathbf{Tr}(\rho Y) \cdot \ket{+i}\bra{+i} - \mathbf{Tr}(\rho Y) \cdot \ket{-i}\bra{-i}.
\label{eq:non-separate-cutting}
\end{align}

\(\mathbf{Tr}(\rho I) \cdot I\) can be replaced by any observable. The total number of subcircuits for one cut is \(6\), which corresponds to the following:
\begin{itemize}
    \item Measuring on the \(Z\) with initial states \(\ket{0}, \ket{1}\)
    \item Measuring on the \(X\) with initial states \(\ket{+}, \ket{-}\)
    \item Measuring on the \(Y\) with initial states \(\ket{+i}, \ket{-i}\).
\end{itemize}

For \(K\) cuts, the total number of subcircuits to run is \(6^K\).

\subsection{Non-Separate Circuit Cutting with the SDO framework}

% The non-separate cut is no longer restricted by the cutting location or the number of cuts. Due to this flexibility, we can choose locations that optimize the number of gates while avoiding excessive cuts, thereby reducing overhead.
Applying the SDO framework to non-separate circuit cutting proceeds similarly to circuit cutting, with two differences.

First, because of the non-separation constraint, we cannot perform biased observable selection for ISDO. Rather than choosing an orthonormal reconstruction basis, each downstream qubit must be initialized in an eigenstate of the upstream measurement observable. Therefore, there is no room for selection.
% For example, if the upstream qubit is measured in the \(X\) basis, the downstream qubit must be initialized in the corresponding eigenstates \(\{\ket{+},\ket{-}\}\).

Second, while doing MSDO, there is no additional overhead for non-separate circuit cutting because each observable requires two executions per cut. For example, if we want to perform MSDO on \(X\), we can replace measurements on the \(X\) observable with projectors \(\ket{+} \bra{+}\) and \(\ket{-} \bra{-}\) with the following:

\begin{itemize}
    \item Measuring on the \(\ket{+} \bra{+}\) with the initial state \(\ket{+}\)
    \item Measuring on the \(\ket{-} \bra{-}\) with the initial state \(\ket{-}\).
\end{itemize}

\begin{table}[htp]
\captionsetup{font=small}
\centering
\renewcommand{\arraystretch}{1.5}
\setlength{\tabcolsep}{5pt}
\begin{tabular}{c c}
\toprule
\multicolumn{2}{c}{\textbf{Uncut QAOA Circuit}} \\
\midrule
\multicolumn{2}{c}{\includegraphics[width=0.45\linewidth]{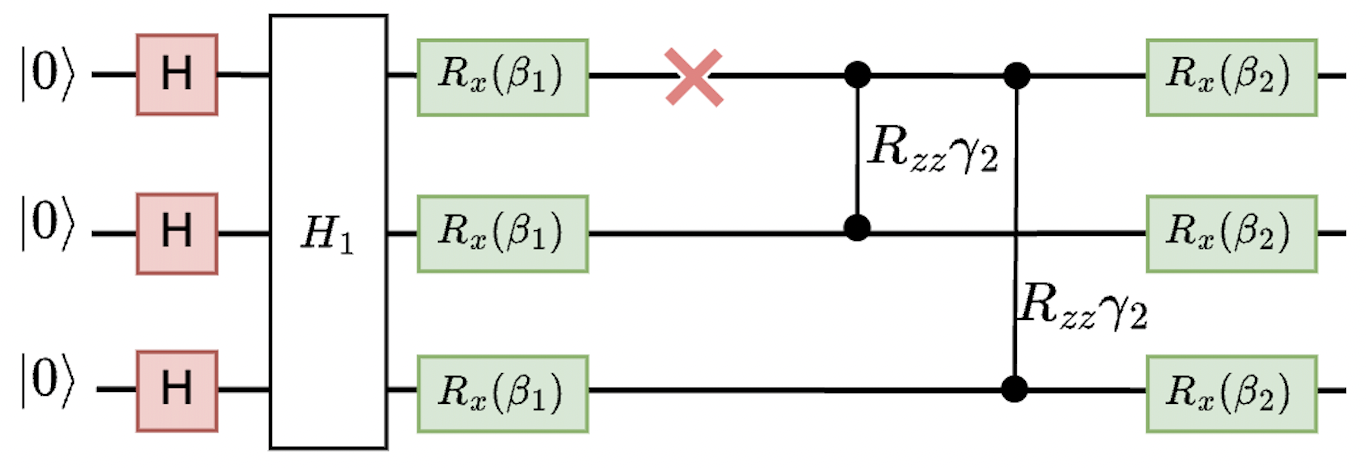}} \\
\midrule
\textbf{Z Measurement $\ket{0}$} & \textbf{Z Measurement $\ket{1}$} \\
\midrule
\includegraphics[width=0.45\linewidth]{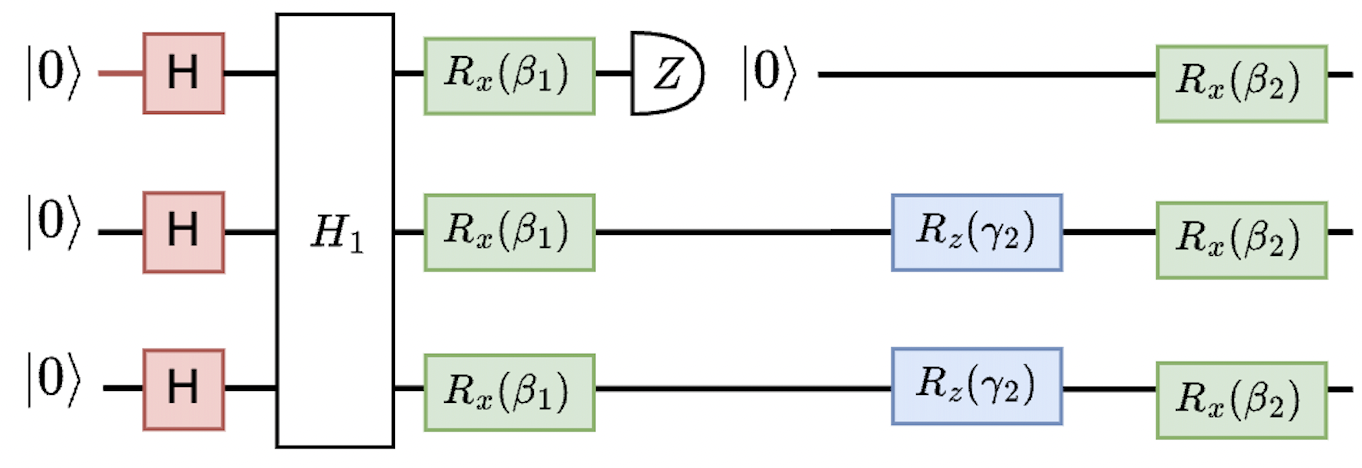} &
\includegraphics[width=0.45\linewidth]{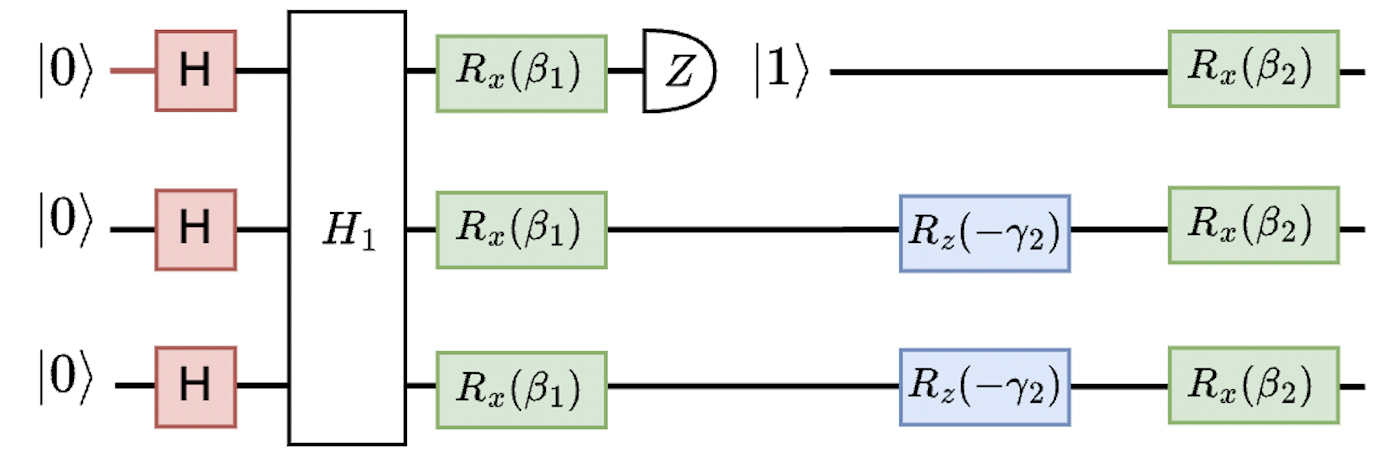} \\
\midrule
\textbf{X Measurement $\ket{+}$ $\ket{-}$} & \textbf{Y Measurement $\ket{+i}$ $\ket{-i}$} \\
\midrule
\includegraphics[width=0.45\linewidth]{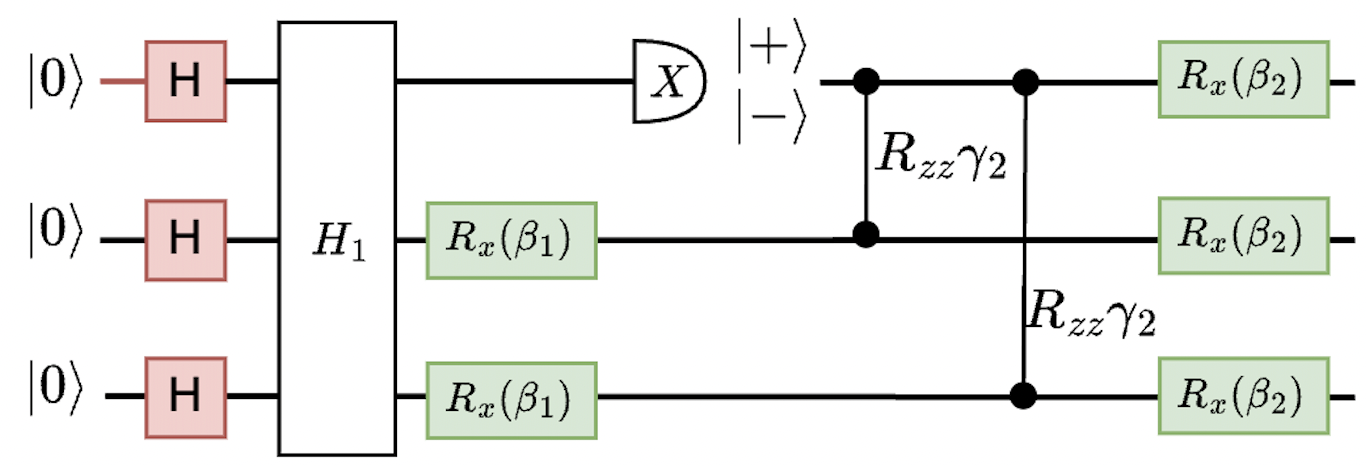} &
\includegraphics[width=0.45\linewidth]{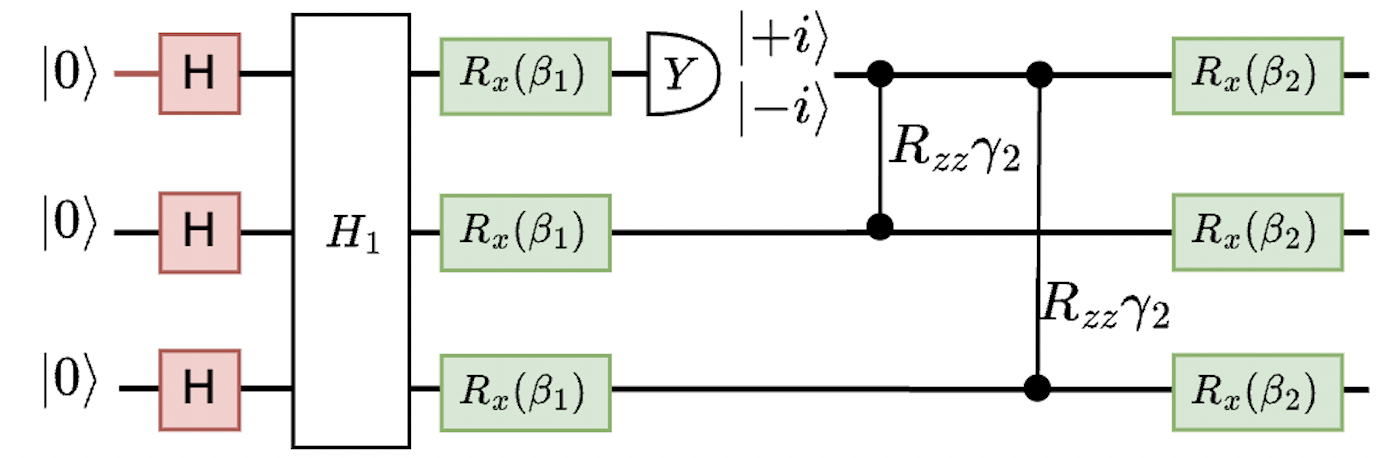} \\
\bottomrule
\end{tabular}
\caption{Example of applying \textit{non-separate circuit cutting} for SDO. The top shows the uncut QAOA circuit, and below are the six subcircuits (the figures in the bottom row each include two subcircuits). \(H_1\) represents the Hamiltonian layer of the QAOA circuit, consistent with the second layer, since both include two RZZ gates. }
\label{tab: Non-Seperate Cutting}
\end{table}

We also use QAOA as an example, as shown in the Table ~\ref{tab: Non-Seperate Cutting}. By cutting at the first qubit, the upstream RX gate can be removed if we measure in the X observable basis. Similarly, downstream, two RZZ gates can be optimized. Among the four reconstruction terms, three benefit from the SDO framework.

% \subsection{Biased Observable Selection for Non-Separate Circuit Cutting}
% \label{subsec: Biased Observable Selection for Non-Separate Circuit Cutting}

% Similar to Circuit Cutting, for Non-Separate Circuit Cutting, we can also calculate the term \(\mathbf{Tr}(I\rho)I\) with other obserables. However, unlike circuit cutting, we do not replace \(\ket{-}, \ket{-i}\) with \(\ket{+}, \ket{+i}\), and \(I\). This is because, in circuit cutting, such replacements can reduce the number of subcircuits, whereas in NSCC, the number of subcircuits actually increases.

% To determine the best observable, we can the method in ~\ref{subsec: Evaluation of the Number of Optimized Gates}. To determine where to cut, we identify a point that maximize the number of optimized gates once a time and repeat greedily to identify as many cut points as possible.

\begin{figure*}[ht]
    \centering
    % Subfigure 1: QAOA 10
    \begin{subfigure}[b]{0.30\textwidth}
        \centering
        \includegraphics[width=\textwidth]{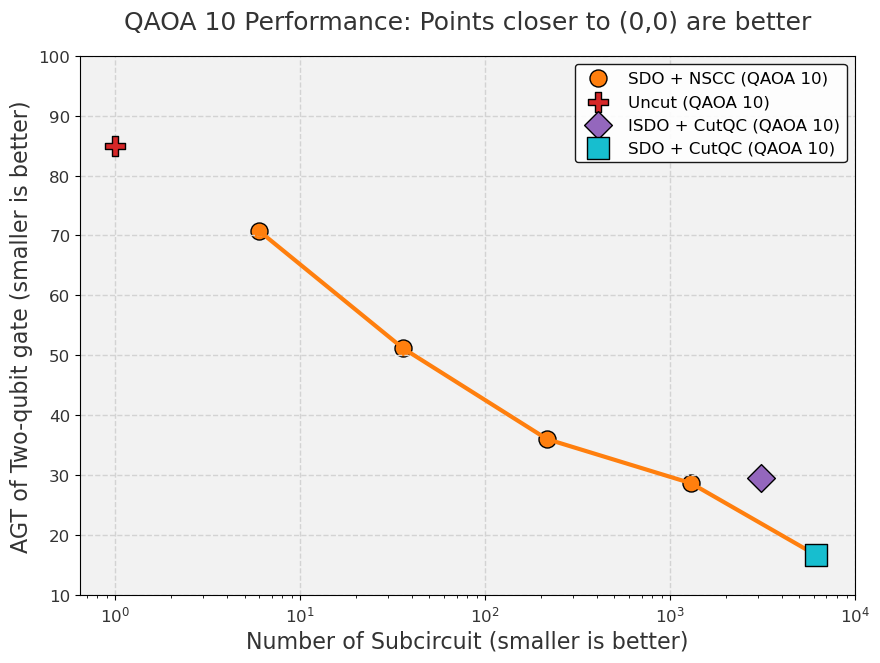} % Replace with actual figure filename
        \caption{QAOA 10 qubits}
        \label{fig:qaoa10}
    \end{subfigure}
    \hfill
    % Subfigure 2: QAOA 20
    \begin{subfigure}[b]{0.30\textwidth}
        \centering
        \includegraphics[width=\textwidth]{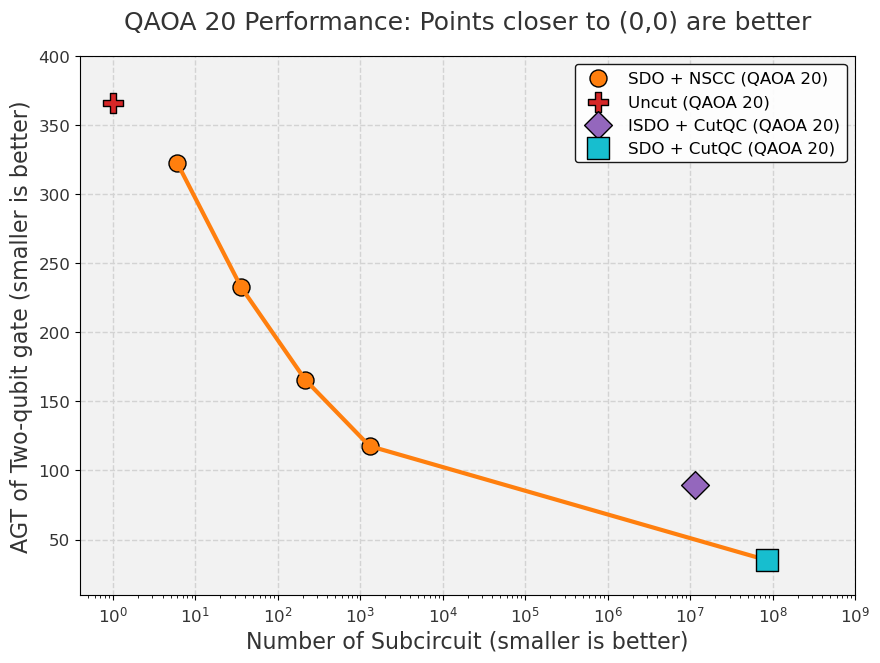} % Replace with actual figure filename
        \caption{QAOA 20 qubits}
        \label{fig:qaoa20}
    \end{subfigure}
    \hfill
    % Subfigure 3: QAOA 30
    \begin{subfigure}[b]{0.30\textwidth}
        \centering
        \includegraphics[width=\textwidth]{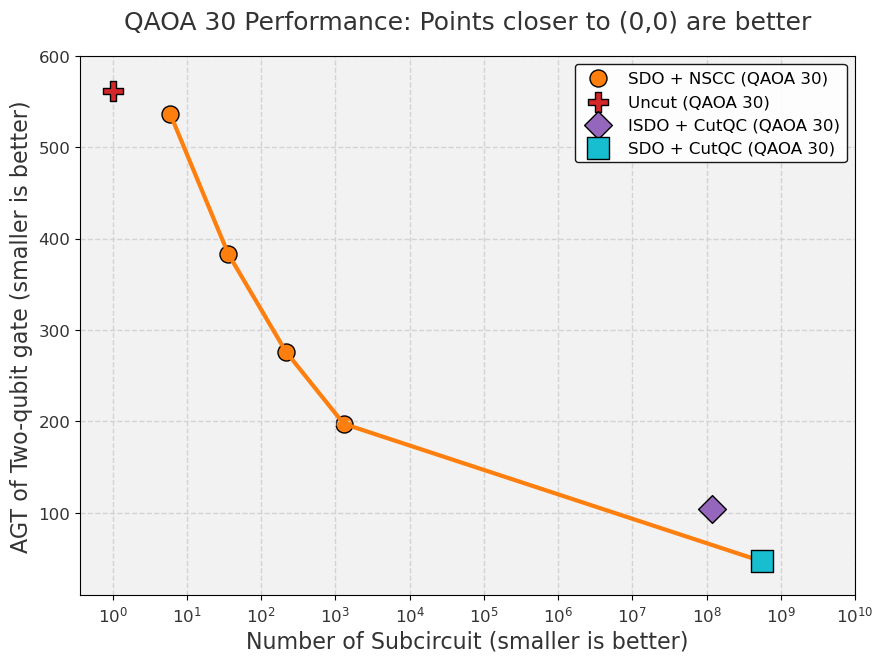} % Replace with actual figure filename
        \caption{QAOA 30 qubits}
        \label{fig:qaoa30}
    \end{subfigure}
    \caption{AGT of two-qubit gates versus the number of subcircuits for different QAOA configurations. The best situation is for each circuit to have fewer subcircuits and a lower AGT. In the case of uncut (red star) and cut (blue square), there are either too many gates or too many subcircuits. The non-separate circuit cutting (NSCC) (orange round) provides an intermediate solution.}
    \label{fig:three_qaoa}
\end{figure*}

\section{Experiments}
\label{sec: Experiment}
In the experiments we first introduce a metric to evaluate the number of gates for reconstructing an outcome. We next present experiments under a noise simulator. We then provide gate information for larger circuits.
\subsection{Average Gate per Term }
For the SDO framework, the number of gate reductions is not as straightforward as directly optimizing gates in the circuit. Instead, it involves optimizing gates in subcircuits and performing calculations to reconstruct the final result. To estimate the gate count, we calculate the average number of gates per term. 
% In total, there are \(4^K\) terms for \(K\) cuts, with each term contributed by subcircuits from all fragments. 
The average is computed within each fragment, and the results across fragments are summed. We refer to this metric as the ``average gate per term" (AGT).

For example, in Table \ref{tab: Circuit Cutting SDO}, assume that we calculate AGT only for logical two-qubit gates. The original Uncut circuit has two RZZ gates; thus AGT is 2. For the cut circuit, the first term \(\mathbf{Tr}(I\rho)I\) has 0 RZZ gates, so the first term contributes 0. The second term is \(\mathbf{Tr}(X\rho)X\), where we examine the \(X\) measurement for upstream, \(X\) initialization with \(\ket{+}\), and \(Z\) initialization with \(\ket{0}\) and \(\ket{1}\) for downstream. The resulting gate count is \(1 + (2 \times 1 + 0 + 0)/2 = 2\). Similarly, the \(\mathbf{Tr}(Y\rho)Y\) term contributes 2, and the \(\mathbf{Tr}(Z\rho)Z\) term contributes 0. Therefore, the AGT in this example is \((0+2+2+0)/4 =1\). 

% In the example of Non-Separate Cutting (Table \ref{tab: Non-Seperate Cutting}), the circuit has two layers, each containing two \(R_{ZZ}\) gates. The AGT of two-qubit gates is calculated as \((2 + 2 + 4 + 4)/4 = 3\). 
% Note that AGT is always weaker in evaluation compared to the number of gates, as it is only an average.

\subsection{Results in the Noise Simulator}
\textbf{Experiments Setup:} Our experiments are conducted using the fake backend Fake127QPulseV1. All circuits are transpiled into physically executable circuits. We tested on the Quantum approximate optimization algorithm (QAOA), quantum Fourier transform (QFT), and Bernstein--Vazirani (BV) circuits. For QAOA, we generated circuits using a random graph with a density of 0.3. We used single-layer (\(p\)=1) QAOA for circuit cutting because a multilayer QAOA cuts solution can be extended from a single layer. For circuit cutting, we used the mixed-integer programming (MIP) model to find the cutting locations. The restrictions for the maximum width were set to half the number of qubits for QAOA and 4 for QFT. For non-separate circuit cutting, we used the greedy algorithm described in Section~\ref{subsec: Biased Observable Selection for Non-Separate Circuit Cutting} to find the cutting locations. Our primary focus is on reducing the number of two-qubit gates in the IBM 127-qubit quantum device. Thus, we calculated the AGT of two-qubit gate and the fidelity for each case. Each case was repeated five times to calculate the average and error bars. The experiment is shown in Table~\ref{tab: SDO for Circuit Cutting} and Table~\ref{tab: SDO for NSCC}.

\begin{table}[h!]
\captionsetup{font=small}
    \centering
    \renewcommand{\arraystretch}{1.2}
    \setlength{\tabcolsep}{2pt}
    \footnotesize % Set font size to footnotesize for better readability and compactness
    \begin{tabular}{c c c c}
        \toprule
        \textbf{Circuit Type} & \textbf{Uncut} & \textbf{Cut} & \textbf{Cut + SDO} \\ \midrule
        \multirow{2}{*}{\parbox{1.5cm}{\centering QAOA 6 (1) \\ \footnotesize [6 cuts]}} 
        & $0.9775 \pm 1.6 \, \epsilon$ & $0.9736 \pm 1.4 \, \epsilon$ & $\textbf{0.9895} \pm 1.8 \, \epsilon$ \\ \cmidrule{2-4}
        & AGT: 33.0  & AGT: 24.0 & AGT: 8.8 \\ \midrule
        \multirow{2}{*}{\parbox{1.5cm}{\centering QAOA 8 (1) \\ \footnotesize [6 cuts]}} 
        & $0.9441 \pm 4.7 \, \epsilon$ & $0.9364 \pm 1.1 \, \epsilon$ & $\textbf{0.9565} \pm 0.4 \, \epsilon$ \\ \cmidrule{2-4}
        & AGT: 53.8  & AGT: 40.15 & AGT: 15.2 \\ \midrule
        \multirow{2}{*}{\parbox{1.5cm}{\centering QAOA 10 (1) \\ \footnotesize [9 cuts]}} 
        & $0.9229 \pm 3.5 \, \epsilon$ & $0.9361 \pm 7.8 \, \epsilon$ & $\textbf{0.9526} \pm 0.9 \, \epsilon$ \\ \cmidrule{2-4}
        & AGT: 86.8 & AGT: 64.6 & AGT: 23.7 \\ \midrule
        \multirow{2}{*}{\parbox{1.5cm}{\centering QFT 6 \\ \footnotesize [6 cuts]}} 
        & $0.9376 \pm 3.7 \, \epsilon$ & $0.9349 \pm 9.0 \, \epsilon$ & $\textbf{0.9535} \pm 1.6 \, \epsilon$ \\ \cmidrule{2-4}
        & AGT: 54.6 & AGT: 45.3 & AGT: 14.0 \\ \midrule
        \multirow{2}{*}{\parbox{1.5cm}{\centering BV 10 \\ \footnotesize [3 cuts]}} 
        & $0.6654 \pm 56.0 \, \epsilon$ & $0.7634 \pm 1.9 \, \epsilon$ & $\textbf{0.7794} \pm 18.0 \, \epsilon$ \\ \cmidrule{2-4}
        & AGT: 31.0 & AGT: 20.0 &  AGT: 8.2 \\
         \bottomrule
    \end{tabular}
    
    % GAIL == my machine jammed in the above and deleted text. I thin I recovered it but do check.
    \caption{\textit{State-dependent optimization} framework on circuit cutting: This table presents results for QAOA, QFT, and BV circuits. We calculate the fidelity and AGT, provide error bars ($\epsilon = 10^{-3}$), and highlight the highest fidelity values. Each benchmark is labeled as [Circuit Name Qubits (Layers)]. For example, QAOA 6 (1) means a 6-qubit QAOA circuit with 1 layer.}
    \label{tab: SDO for Circuit Cutting}
\end{table}

\begin{table}[h!]
\captionsetup{font=small}
    \centering
    \renewcommand{\arraystretch}{1.2}
    \setlength{\tabcolsep}{2pt}
    \footnotesize % Set font size to footnotesize for better readability and compactness
    \begin{tabular}{c c c c}
        \toprule
        \textbf{Circuit Type} & \textbf{Uncut} & \textbf{1 Cut + SDO} & \textbf{2 Cuts + SDO}  \\ \midrule
        \multirow{2}{*}{QAOA 6 (2)} 
        & $0.9780 \pm 1.5 \, \epsilon$ & $0.9828 \pm 2.2 \, \epsilon$ & $\textbf{0.9862} \pm 1.5 \, \epsilon$ \\ \cmidrule{2-4}
        & AGT: 176.8  & AGT: 166.8  & AGT: 132.0  \\ \midrule
        \multirow{2}{*}{QAOA 8 (2)} 
        & $0.9675 \pm 3.3 \, \epsilon$ & $0.9713 \pm 2.1 \, \epsilon$ & $\textbf{0.9786} \pm 1.2 \, \epsilon$ \\ \cmidrule{2-4}
        & AGT: 183  & AGT: 165.9  & AGT: 139.9 \\ \midrule
        \multirow{2}{*}{QAOA 10 (2)} 
        & $0.8729 \pm 8.6 \, \epsilon$ & $0.8942 \pm 8.0 \, \epsilon$ & $\textbf{0.9085} \pm 3.7 \, \epsilon$ \\ \cmidrule{2-4}
        & AGT: 290.2  & AGT: 271.4 & AGT: 219.0 \\ \midrule
        \multirow{2}{*}{QFT 10 } 
        & $0.8967 \pm 5.9  \, \epsilon$ & $0.9264 \pm 12.3\, \epsilon$ & $\textbf{0.9401} \pm 2.7 \, \epsilon$ \\ \cmidrule{2-4}
        & AGT: 197.4 & AGT: 180.5 & AGT: 130.8 \\ 
        % \midrule
        %  \multicolumn{1}{c}{Type Qubits (Layers)} & \multicolumn{3}{c}{$\epsilon = 10^{-3}$} \\ 
         \bottomrule
    \end{tabular}
    \caption{\textit{State-dependent optimization}  framework on \textit{non-separate circuit cutting}. This table presents results for QAOA and QFT. For the BV circuit, the situation is similar to circuit cutting, since introducing any cut would separate the circuit. We calculate the fidelity and AGT, provide error bars ($\epsilon = 10^{-3}$), and highlight the highest fidelity values. Each benchmark is labeled as [Circuit Name Qubits (Layers)]. }
    \label{tab: SDO for NSCC}
\end{table}

\textbf{Result Summary:} Applying the SDO framework on circuit cutting can reduce average AGT by 73.1\% and improve fidelity by an average of 4\%, up to 14\%. Without the SDO framework, circuit cutting may reduce fidelity because ofadditional measurement errors and qubit error. In contrast, integrating our SDO with circuit cutting effectively mitigates these errors and consistently improves fidelity.

Applying the SDO framework on non-separate circuit cutting can reduce average AGT by 16.87\% and improves fidelity by an average of 2.7\%, up to 4.0\%. The limited improvement is due to fewer cuts being introduced compared with circuit cutting and the greater difficulty of mapping qubits compared with the smaller subcircuits produced by circuit cutting.

% Overall, we observed that using SDO always outperforms not using it in both circuit cutting and Non-Separate Circuit Cutting.

\subsection{Subcircuit Scaling and AGT}
Based on the experiments shown in Figure~\ref{fig:three_qaoa}, we aim to demonstrate reasonable deductions for larger circuits. Using QAOA (\(p\)=1) as an example, we generate circuits with 10, 20, and 30 nodes, each with a density of 0.3. For circuit cutting, we restrict the maximum subcircuit size to 10 and employ an MIP model to find solutions. Cutting these circuits leads to an exponential scaling in the number of subcircuits. Instead of relying solely on circuit cutting, however, non-separate circuit cutting offers an approach that reduces AGT by running fewer subcircuits.

\section{Conclusion}
\label{sec: Conclusion}
In this paper we demonstrate that the circuit optimization technique ISDO is particularly well suited for wire cutting. To enhance its robustness, we introduce MSDO and a biased observable selection strategy, combining all together into a unified state-dependent optimization (SDO) framework. We test the SDO framework on circuit cutting, showing steady fidelity improvement. Recognizing the limitations of circuit cutting, including restrictions on cutting locations and the number of cuts, we propose non-separate circuit cutting, which utilizes wire cutting without dividing the circuit into isolated subcircuits. Our experiments show that the SDO framework on non-separate circuit cutting also achieves fidelity improvement, providing an effective alternative for optimizing quantum circuits.

\section*{Acknowledgment}
This work was supported by grants from NSF 2216923 and  2117439. 
This material is based upon work supported by the U.S. Department of Energy,
Office of Science, under contract number DE-AC02-06CH11357. 
% Research supported by Q-NEXT, one of the U.S.~Department of Energy Office of
% Science National Quantum Information Science Research Centers and the Office of
% Advanced Scientific Computing Research, Accelerated Research for Quantum
% Computing program.
% Not sure if we want per-research acknowledgments: 
Larson was supported by Q-NEXT, one of the U.S.~Department of Energy Office of Science National Quantum Information Science Research Centers. 
Liu and Hovland were supported by the  U.S.~Department of Energy, Office of Science, Office of Advanced Scientific Computing Research, Accelerated Research for Quantum Computing program

\bibliographystyle{IEEEtran}
\bibliography{IEEEabrv,cite}

\vfill
\framebox{\parbox{.90\linewidth}{\scriptsize The submitted manuscript has been
created by UChicago Argonne, LLC, Operator of Argonne National Laboratory
(``Argonne''). Argonne, a U.S.\ Department of Energy Office of Science
laboratory, is operated under Contract No.\ DE-AC02-06CH11357.  The U.S.\
Government retains for itself, and others acting on its behalf, a paid-up
nonexclusive, irrevocable worldwide license in said article to reproduce,
prepare derivative works, distribute copies to the public, and perform publicly
and display publicly, by or on behalf of the Government.  The Department of
Energy will provide public access to these results of federally sponsored
research in accordance with the DOE Public Access Plan
\url{http://energy.gov/downloads/doe-public-access-plan}.}}
\end{document}